\documentclass{svproc}
\usepackage{url}

\usepackage{graphicx}

\begin{document}
\mainmatter              
\title{Multi-modal Summarization in Model-Based Engineering: Automotive Software Development Case Study}
\titlerunning{Multi-modal Summarization in Model-Based Engineering}  
\author{Nenad Petrovic\inst{1} \and Yurui Zhang\inst{1} \and
 Moaad Maaroufi\inst{1} \and  Kuo-Yi Chao \inst{1} \and Lukasz Mazur\inst{1} \and Fengjunjie Pan\inst{1} \and Vahid Zolfaghari\inst{1} \and Alois Knoll\inst{1}
}
\authorrunning{Nenad Petrovic et al.} 
%
\tocauthor{First Author, Second Author, Third Author and Fourth Author}
\institute{Technical University of Munich, Robotics, Artificial Intelligence and Real-Time Systems, Munich, Germany,\\
\email{nenad.petrovic@tum.de, yurui.zhang@tum.de, moaad.maaroufi@tum.de, kuoyi.chao@tum.de, lukasz.mazur@tum.de, f.pan@tum.de, v.zolfaghari@tum.de, knoll@in.tum.de}
}

\maketitle              

\begin{abstract}
Multimodal summarization integrating information from diverse data modalities presents a promising solution to aid the understanding of information within various processes. However, the application and advantages of multimodal summarization have not received much attention in model-based engineering (MBE), where it has become a cornerstone in the design and development of complex systems, leveraging formal models to improve understanding, validation and automation throughout the engineering lifecycle. UML and EMF diagrams in model-based engineering contain a large amount of multimodal information and intricate relational data. Hence, our study explores the application of multimodal large language models within the domain of model-based engineering to evaluate their capacity for understanding and identifying relationships, features, and functionalities embedded in UML and EMF diagrams. We aim to demonstrate the transformative potential benefits and limitations of multimodal summarization in improving productivity and accuracy in MBE practices. The proposed approach is evaluated within the context of automotive software development, while many promising state-of-art models were taken into account.
\end{abstract}
\section{Introduction}
Multimodal Large Language Models (MLLMs)\cite{zhang2024mmllmsrecentadvancesmultimodal} represent a significant advancement in artificial intelligence, extending the capabilities of traditional language models to process and generate data across multiple modalities, such as text, images, audio, and video. Unlike conventional large language models (LLMs)\cite{zhao2024surveylargelanguagemodels} that focus solely on textual information, MLLMs are designed to integrate and interpret diverse forms of data, enabling them to address complex, real-world challenges\cite{liang2024comprehensivesurveyguidemultimodal} where information is often transferred through a combination of modalities. Consequently, the aviation and automotive industries are increasingly leveraging MLLMs to address complex real-world challenges and use cases in industrial design. In particular, Model-Based Engineering (MBE) demands that MLLMs accurately comprehend and handle systematic approaches for complex tasks such as requirements management, system analysis, design, validation, and verification\cite{10741141}\cite{llmmde} which are mostly presented by text descriptions and visual diagrams of Eclipse Modeling Framework (EMF)\cite{Steinberg2008PearsonEducation} and Unified Modeling Language (UML)\cite{Pilone2005OReillyMedia} as in the automotive domain example of Centralized Car Server Metamodel from \cite{llmmde}. Enabling MLLMs to accurately analyze class-to-class relationships, as well as the properties and functionalities of classes from metamodeling diagrams in UML and EMF, has become a key research focus in the industrial domains\cite{geipel2024towards}.

Although research on the capability of MLLMs in analyzing MBE diagrams remains rare attention (especially in automotive), the past three years have witnessed significant studies on MLLMs or Vision-Language Models (VLMs)\cite{Zhang_2024_WACV}\cite{che2023enhancing} in domains such as multimodal summarization\cite{jangra2023survey}, multimodal Chain-of-Thought\cite{zhang2023multimodal}, textbook question answering\cite{lu2022learn}\cite{tan2024retrieval}\cite{ma2022weakly} and diagram-based question answering\cite{wang2022computer}\cite{wang2024cog}, etc. However, a lot of information beneficial for maintainability and updates of older vehicles resides in different types of diagrams, which needs lots of time and effort to be analyzed by experts. Therefore, this study primarily explores whether existing techniques can be applied to the analysis of MBSE diagrams, with focus on automotive industry usage. Additionally, it examines the current challenges and limitations in this emerging field and presents the workflow aiming automated development of automotive software as one of the outcomes. Numerous state-of-the-art models are compared side-by-side, while the most promising one was used for proof-of-concept implementation shown towards the end of the paper.

The rest of the paper has the following structure. Next section provides overview of related works, covering both MLLMs and approaches leveraging them for diagram prompting. Additionally, this section also gives tabular summary of relevant state-of-art models, considering their parameter number and usage costs among other factors. The third section describes our experiment from automotive domain which was used for comparative evaluation of the selected models. The fourth section focuses on adoption of MLLMs within automotive software development toolchain. The fifth section shows results of evaluation for the selected models. Finally, the conclusion summarizes the main contributions achieved and aspects observed during evaluation. 
\section{Related Works}

\subsection{Multimodal Large Language Models}
Since the release of ChatGPT as a LLM in December 2022, the field of MLLMs has experienced explosive growth. Over the past two years, various AI technology companies and academic research institutions have developed and publicly released their own MLLM models. Notable examples include OpenAI's GPT-4 series\cite{openai2024gpt4technicalreport}, Google's Gemini series\cite{geminiteam2024gemini15unlockingmultimodal}, Meta's Llama series\cite{grattafiori2024llama3herdmodels}, Anthropic's Claude 3\cite{TheClaude3}, Mistral's Pixtral\cite{agrawal2024pixtral12b}, xAI's Grok\cite{xAIGrok}, and others\cite{wang2024emu3nexttokenpredictionneed}\cite{xue2024xgenmmblip3familyopen}\cite{Fuyu8b}\cite{jin2309unified}\cite{peng2023kosmos}\cite{internvl2}.
The methods for utilizing these MLLMs are highly diverse. Some offer interactive user interfaces or API access, others provide only limited test access. While some models are open-source, others are entirely closed-source. In certain cases, models are limited to demos, making it impossible for external users to directly test or use them. Therefore, Table 1 presents an clear overview of currently available models that external users can directly access. It includes details such as model size, release date, development organization, access usage methods, open-source status, and whether the model requires payment.

Recent researches\cite{conrardy2024image}\cite{rossi2024importance} have explored MLLMs to generate sample UML diagrams from drawings. These models translate drawn visual elements into structured representations, enabling automated and efficient diagram-to-model conversions. This highlights the potential of MLLMs in bridging visual and formal representations, particularly in software engineering.
\subsection{Diagram prompting and pre-processing}

Due to MLLM development and limitations in hardware resources, many research institutions are unable to leverage MLLMs for complex diagram image processing. Therefore, early research on diagram understanding in geometry problem-solving focused on integrating visual and textual information to enhance reasoning capabilities. 
The G-ALIGNER model\cite{seo2014diagram} proposed in 2014 introduced a method for diagram understanding by combining visual element detection with textual alignment through submodular optimization, enabling accurate identification and alignment of geometric primitives with corresponding textual descriptions. 

Building on this foundation, the Weakly Supervised Learning for Textbook Question Answering (WSTQ) framework\cite{ma2022weakly} in 2022 utilized weak supervision from text retrieval and object detection to develop text matching and relation detection tasks, significantly improving accuracy on the CK12-QA\cite{ma2023xtqa} and AI2D\cite{hiippala2021ai2d} datasets through multitask learning. 
In the same year, PGDP-Net\cite{zhang2022plane} was introduced as an end-to-end solution for plane geometry diagram parsing, employing a modified instance segmentation method for geometric primitive extraction and a Graph Neural Network (GNN)\cite{wu2020comprehensive} for relation parsing, supported by the comprehensive PGDP5K dataset. 
In 2023, the Multimodal Chain-of-Thought (MCoT) framework\cite{zhang2023multimodal} proposed a two-stage reasoning approach that separates rationale generation from answer inference, effectively mitigating hallucinations and achieving state-of-the-art performance on multimodal reasoning tasks such as ScienceQA\cite{saikh2022scienceqa} and A-OKVQA\cite{schwenk2022okvqa}. 

Continuing this progression, the CoG-DQA framework\cite{wang2024cog} introduced in 2024 leverages Large Language Models (LLMs) to guide Diagram Parsing Tools (DPTs) through a chain-of-guiding mechanism, integrating visual parsing with domain-specific knowledge to enhance diagram question answering tasks\cite{wang2021csdqa}\cite{hiippala2021ai2d}. 
Most recently, the DiagramQG\cite{zhang2024diagramqg} dataset and its Hierarchical Knowledge Integration framework (HKI-DQG) were developed to generate concept-focused educational questions from diagrams, utilizing advanced vision-language models to surpass existing methods in educational question generation tasks. 

Collectively, these studies demonstrate a clear trajectory of progress in multimodal reasoning and diagram understanding, highlighting the growing effectiveness of integrating visual and textual information in complex reasoning tasks.
\begin{table}[htbp]
\caption{A summary of commonly used MLLMs.}
\centering
\begin{tabular}{|p{2.1cm}|p{2.0cm}|p{2.2cm}|p{2.3cm}|p{2.4cm}|p{2.2cm}|}
\hline
\textbf{Model} & \textbf{Release Date} & \textbf{Organization} & \textbf{Parameter Size (B)} & \textbf{Access Usage Methods} & \textbf{Cost} \\
\hline
Grok-2 & Dec-2024 & xAI & $>$314 & Website \& API & Paid \\
Emu3 & Sep-2024 & BAAI & 8 & Open Source (Code \& Model) & Free \\
Llama 3.2 & Sep-2024 & Meta & 11 / 90 & Open Source (Code \& Model) & Free \\
Pixtral-12B & Sep-2024 & Mistral & 12 & Website \& API & Paid \\
Llama 3 & Jul-2024 & Meta & 8 / 70 / 405 & Open Source (Code \& Model) & Free \\
InternVL2 & July-2024 & OpenGVLab & 8 & Open Source (Code \& Model) & Free \\
xGen-MM (BLIP-3) & Aug-2024 & Salesforce AI & 4 & Open Source (Code \& Model) & Free \\
Chameleon & May-2024 & Meta & 7 / 34 & Open Source (Code \& Model) & Free \\
GPT-4o & May-2024 & OpenAI & unknown & Website \& API & Paid \\
Claude 3 & Mar-2024 & Anthropic & $>$175 & Website \& API & Paid \\
Grok-1 & Mar-2024 & xAI & 314 & Open Source (Code \& Model) & Free \\
Gemini 1.5 & Feb-2024 & Google & unknown & Website \& API & Paid \\
Fuyu-8B & Oct-2023 & Adept & 8 & Open Source (Code \& Model) & Free \\
PaLI-3 & Oct-2023 & Google DeepMind & 2 / 3 / 5 & Open Source (Code \& Model) & Free \\
GPT-4V & Sep-2023 & OpenAI & unknown & Website \& API & Paid \\
LaVIT & Sep-2023 & Peking University & 7 & Open Source (Code \& Model) & Free \\
Emu1 & Jul-2023 & BAAI & 14 & Open Source (Code \& Model) & Free \\
UnIVAL & Jul-2023 & Sorbonne University & 0.25 & Open Source (Code \& Model) & Free \\
KOSMOS-2 & Jun-2023 & Microsoft Research & 7 & Open Source (Code \& Model) & Free \\
GPT-4 & Mar-2023 & OpenAI & unknown & Website \& API & Paid \\
\hline
\end{tabular}
\label{tab:open_llms}
\end{table}

\section{Experiment}

\subsection{Research Questions}
Building upon the exploration of MLLMs and VLMs for diagram recognition, this subsection investigates the recognition capabilities of MLLMs in the automotive manufacturing and autonomous driving industry, specifically focusing on complex UML class diagrams for automotive components.

We propose the following research questions inspired by \cite{conrardy2024image}:

\begin{itemize}
    \item \textbf{RQ1:} Can MLLMs accurately identify all categories of automobile components depicted in a UML class diagram?
    \item \textbf{RQ2:} Are MLLMs capable of recognizing and understanding the functional descriptions of automotive components within the UML class diagram?  
    \item \textbf{RQ3:} Can MLLMs correctly classify automotive components into their appropriate categories based on the relationships and descriptions in the UML class diagram?  
    \item \textbf{RQ4:} Can MLLMs accurately identify and map the relationship chains between automotive components as illustrated in the UML class diagram?
    \item \textbf{RQ5:} Can MLLMs accurately detect the differences between the most similar UML class diagrams?
\end{itemize}
These research questions aim to assess the potential of MLLMs in analyzing and transforming automotive UML diagrams into semantically correct machine-readable formats. The focus is on testing MLLMs' abilities in recognizing structural, relational, and functional information, which is critical for the automotive manufacturing and autonomous driving sectors.
\subsection{Experimental Setup}
To correspond with the research questions mentioned above, we designed five questions based on visually represented model with respect to \cite{llmmde}:
\begin{itemize}
    \item \textbf{Q1:} Given a UML diagram about Centralized Car Server Metamodel, list all classes in this UML diagram
    \item \textbf{Q2:} Given a UML diagram about Centralized Car Server Metamodel, list all properties and functions in processing node class
    \item \textbf{Q3:} Given a UML diagram about Centralized Car Server Metamodel, is FPGA one of the Co-Processor? \textit{(A)} / is the camera sensor? \textit{(B)}.
    \item \textbf{Q4:} Given a UML diagram about Centralized Car Server Metamodel, list all classes on the relation chain between camera and component \textit{(A)} / list all subclasses that processing task class has \textit{(B)}.
    \item \textbf{Q5:} Given \textit{two} UML diagrams about Centralized Car Server Metamodel, what are the differences between these diagrams?
\end{itemize}
In Q5, we remove the GPU, FPGA, and TPU classes displayed in the UML diagram, leaving blank spaces, and assign the MLLMs to identify the differences between these two UML diagrams.

We also defined ground truths (GT) for each of the five different questions to better evaluate the performance of the MLLMs.
\begin{itemize}
    \item \textbf{GT1:} The MLLMs are required to identify the names of all classes as well as the total number of classes.
    \item \textbf{GT2:} The MLLMs are required to accurately list the names and types of attributes and functions within the Processing Node class.
    \item \textbf{GT3:} The MLLMs only need to respond with a "yes" or "no" for question \textit{A} and \textit{B}.  
    \item \textbf{GT4:} The MLLMs need to list the correct four class names in the relationship chain between Camera and Component \textit{(A)} and the three subclasses of the Process Task class \textit{(B)}.
    \item \textbf{GT5:} The MLLMs need to identify that the second UML class diagram, which has been manually modified, lacks the GPU, FPGA and TPU classes.
\end{itemize}

Due to hardware limitations, we evaluated mostly the performance of web-based MLLMs. During the testing, we initiated a new conversation, uploaded model diagram based on \cite{llmmde}, and sequentially asked five questions. The performance of the MLLMs on these five questions was manually assessed based on ground truths.

\section{Evaluation}
In Table \ref{tab:results}, the results of MLLMs on the five questions are presented. For Q1, Q3 and Q4B, almost all MLLMs were able to provide completely correct answers. However, none of the MLLMs detected differences and changes between the two UML diagrams in answering Q5. Furthermore, different models exhibited significant capability differences for Q2. For example, Claude-3.5 provided a completely correct answer, while Gemini-2.0 gave an entirely incorrect response. Other models could answer correctly to varying extents, but many of their answers still contained errors and hallucinations. When asked about the relationship chain between the Camera and Component classes (Q4B), the results from MLLMs showed significant inconsistency. Some models were able to provide a completely accurate relationship chain, while others generated responses that were merely plausible, but ultimately hallucinated. 

Based on this results, we found that current MLLMs can perfectly identify non-complex content in UML diagrams, such as class names, the number of classes, and simple class inheritance relationships. But for more complex questions, such as those involving class attributes and functions as well as relation chains, many models lack the capability to provide correct answers. Most notably, MLLMs entirely lack the ability to recognize differences between two similar but distinct UML diagrams. However, it can be noticed that InternVL2-8B-MPO model which is free to use exhibits capability to answer correctly in case of all the question templates, which is making it suitable for the considered automotive use case. Based on our findings, InternVL2-8B-MPO is very good at processing and reasoning across metamodels, as it manages to understand the hierarchies and attributes and relationships between the components of the metamodel, which is quite impressive with only 8B parameters for the connection layer between the linguistic and visual encoder (visual encoder has 6B and the linguistic encoder has 13B parameters; overall the model has 6+8+13=27B parameters). Additionally, this model is also deployable locally without relying on external providers and services, which is of utmost importance in automotive industry, as companies usually apply policies which prevent sending data to third parties outside the boundaries of the underlying organization.

In summary, Table \ref{tab:results} highlights the strengths and limitations of current MLLMs in answering UML diagram-related questions, providing clear directions for improvement in future research on this topic. 

\begin{table*}[htbp]
\caption{The results of MLLMs performance on 5 questions related to \cite{llmmde}.}
\centering
\resizebox{\textwidth}{!}{%
\begin{tabular}{|l|p{2.5cm}|p{2.5cm}|p{2.5cm}|p{2.5cm}|p{2.5cm}|}
\hline
\textbf{Model}         & \textbf{Q1} & \textbf{Q2} & \textbf{Q3} & \textbf{Q4} & \textbf{Q5} \\ \hline
Grok-2                  & 28/29              & Partially correct with much hallucination        & A\&B correct           & Only B correct           & No correct difference detected        \\ \hline
Pixtral-12B & 29/29 & Mostly correct with lacking 2 attributes and 1 function & A\&B correct & Only B correct & No correct difference detected \\ \hline
Claude-3.5 & 29/29 & Totally correct & A\&B correct & A\&B correct & No correct difference detected \\ \hline
Gemini-2.0 & 28/29 & Totally wrong & A\&B correct & A\&B correct & No correct difference detected \\ \hline
Gemini-1.5 & 29/29 & Partially correct with much hallucination & A\&B correct & A\&B correct & No correct difference detected \\ \hline
GPT-4o & 28/29 & Mostly correct with 1 wrong attribute & A\&B correct & Only B correct & No correct difference detected \\ \hline
GPT-o1 & 29/29 & Mostly correct with 1 wrong attribute and lacking 1 attribute & A\&B correct & A\&B correct & No correct difference detected \\ \hline
GPT-4 & 29/29 & Mostly correct with few hallucination & A\&B correct & A\&B correct & No correct difference detected \\ \hline
GPT-4o-mini & 29/29 & Partially correct with much hallucination & A\&B correct & Only B correct & No correct difference detected \\ \hline
InternVL2-8B-MPO & 29/29 & Totally correct & A\&B correct & A\&B correct & Correct difference detected \\ \hline
Qwen2 VL 7B & 24/29 & Correct Properties, hallucinated functions & A\&B correct & Only B correct & Correct difference detected \\ \hline
\end{tabular}%
}
\label{tab:results}
\end{table*}

\section{Usage in Automotive}

\begin{figure}[htbp]
\centering
\includegraphics[width=\textwidth]{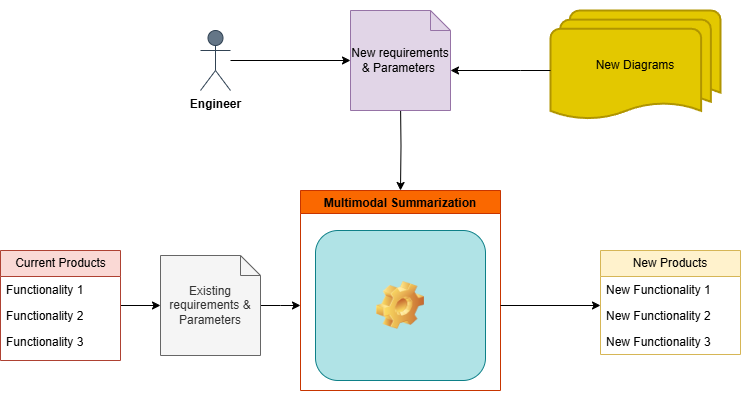}
\caption{MLLM-based diagram prompting for product updates identification.}
\label{fig:workflow}
\end{figure}

In some domains, such as automotive, products may require continuous iterative updates through their life cycle. Each product is associated with a significant amount of requirements, expressed as either parameter information, diagrams (such as UML-alike representations in automotive), tables, and operational instructions. During every product update or modification, a corresponding series of related artifacts (such as documentation, configuration and software code) must be changed accordingly. as illustrated in the Fig. \ref{fig:workflow}. In this process, a large volume of product information needs to be compared and summarized to provide a clear overview of updates. 

\begin{figure}[htbp]
\centering
\includegraphics[width=\textwidth]{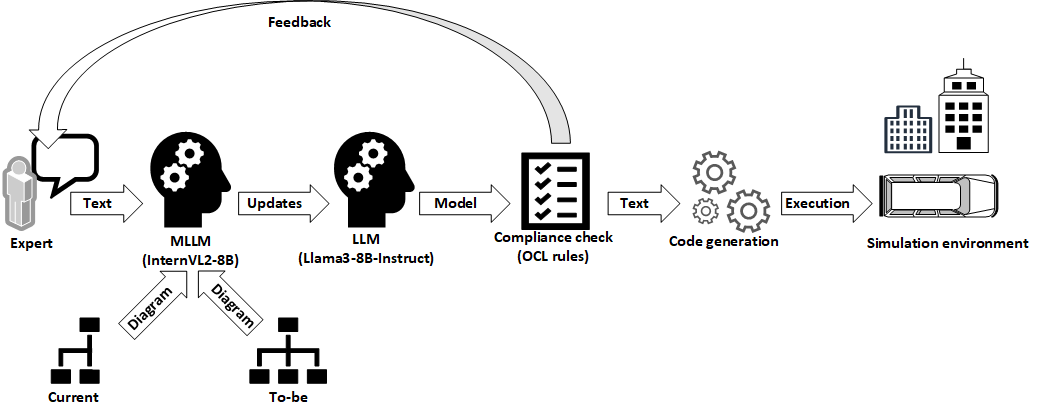}
\caption{The workflow of MLLM-based image prompting for automotive scenarios.}
\label{fig:workflow2}
\end{figure}

In automotive, maintainability and updateability represent challenges due to strict standardization which requires lot of time and efforts, slowing down the innovation. To reduce the labor costs associated with related tasks, we introduce automated approach leveraging MLLMs to perform multimodal information summarization, as depicted in Fig. \ref{fig:workflow2}. In the first steps, user specifies changes of the requirements or new requirements, either as freeform textual description or providing diagram representation of system instance with respect to pre-defined metamodel. Considering the current system representation, updates are identified using MLLM, such as addition of new sensors, actuators or their improvement (camera resolution increase). In the next step, the changes detected as outcome of multimodal summarization are further leveraged as input of LLM-based code generation workflow, targeting CARLA simulation environment. Before the actual code generation, the new requirements are taken into account for updated model instance creation. Moreover, this updated model instance is checked for automotive compliance with respect to given ISO standard, making use of Object Constraint Language (OCL) rules. In case that addition of new requirements leads to system model instance which is not compliant, feedback to the user generated. In that case, user can understand which part of new requirements is not compliant and should be corrected. For model instance creation and feedback generation, we make use of Llama3-8B-Instruct LLM. 

Services for tasks performing image prompting are based on OpenGVLab/ InternVL2-8B-MPO \cite{internvl28b} and deployed as web application relying on Quart \cite{quart} framework for Python. It can also be deployed using lmdeploy \cite{2023lmdeploy} which is a toolkit for compressing, deploying, and serving LLM. Despite that there are variants with more parameters, we select the one with 8B as it is the smallest one which answers to all the questions correctly, while runnable on lower hardware configurations quantized to 4 bit. The deployment was done in Google Colab environment with the lowest Pro subscription plan, as more than 20GB of VRAM was required for execution, which is above the limit of free account. Based on our experience, the lowest configuration offered by Google Colab able to run it was L4 GPU-based, where 22GB of VRAM were occupied. In Table \ref{tab:automotive}, an overview of API and average execution times in seconds for the described L4-based deployment in Google Colab is given. It can be noticed that execution time has order of magnitude of second. Moreover, we can also identify that the detection of differences between diagrams has the slowest execution time, which was expected considering the fact that it includes prompting of both diagrams, so additional processing was needed compared to other cases. 

As the authors mention in \cite{internvl28b}, the model uses a substantial middleware layer between the visual encoder and the LLM unlike than typical lightweight connectors most MLLMs use. Furthermore, the MPO (Mixed Preference Optimization) training which mixes  three losses (preference ranking, quality checks, and generation guidance) allows to enhance the model's reasoning capabilities and reduce hallucinations.

\begin{table}[htbp]
\caption{Automotive diagram prompting service REST API and execution times.}
\centering
\begin{tabular}{|p{2.8cm}|p{2.8cm}| p{3.5cm}|p{2.2cm}|}
\hline
\textbf{Endpoint} & \textbf{Parameters} & \textbf{Description} & \textbf{Execution time [s]}\\
\hline
/extractSensors & Diagram & Returns the list of all sensors from the diagrams & 2.86 \\
/extractActuators &  Diagram & Returns the list of all actuators from the diagram & 2.64 \\
/extractElement \newline Properties & Diagram \newline Element name & Returns the list of properties and values for given element (such as sensor or actuator) & 3.08 \\
/detectDifferences & Current diagram \newline New diagram & Returns the list of differences between two diagrams, considering sensors, actuators, their properties, interfaces and parameter values & 9.13 \\

\hline
\end{tabular}
\label{tab:automotive}
\end{table}

\section{Conclusion}
In this work, we considered some of the promising MLLM solutions currently available,  together with the frameworks and techniques related to the processing of scientific diagrams. Subsequently, we designed five types of research questions for UML diagrams and proposed five example questions along with their corresponding ground truths based on the UML class diagram we have. Using these questions, we tested the web-based MLLMs currently available and evaluated their performance. On the other side, we also show a proof-of-concept implementation of the MLLM-based web service relying on the Intern-VL model.

Future work should be drawn attention to enhancing MLLMs' ability to process set of large and complex UML diagrams describing singular system, particularly in understanding class attributes and functions for purpose of complex scenarios targeting code generation in automotive domain, focusing on hardware abstraction aspects. Meanwhile, improving models' capacity to detect and interpret subtle differences between similar UML diagrams is a critical direction. In addition, developing new benchmarks and datasets that target these challenges in automotive will also be essential to driving progress in this area.

\section{Acknowledgment}
This work has received funding from the European Chips Joint Undertaking under Framework Partnership Agreement No 101139789 (HAL4SDV) including the national funding from the German Federal Ministry of Education and Research (BMBF) under grant number 16MEE00471K. The responsibility for the content of this publication lies with the authors.
%
%


\begin{thebibliography}{6}
%

\bibitem{zhang2024mmllmsrecentadvancesmultimodal}
Zhang, D., Yu, Y., Dong, J., Li, C., Su, D., Chu, C., Yu, D.: MM-LLMs: Recent Advances in MultiModal Large Language Models. arXiv preprint arXiv:2401.13601 (2024). https://arxiv.org/abs/2401.13601

\bibitem{zhao2024surveylargelanguagemodels}
Zhao, W.X., Zhou, K., Li, J., Tang, T., Wang, X., Hou, Y., et al.: A Survey of Large Language Models. arXiv preprint arXiv:2303.18223 (2024). https://arxiv.org/abs/2303.18223

\bibitem{liang2024comprehensivesurveyguidemultimodal}
Liang, C.X., Tian, P., Yin, C.H., et al.: A Comprehensive Survey and Guide to Multimodal Large Language Models in Vision-Language Tasks. arXiv preprint arXiv:2411.06284 (2024). https://arxiv.org/abs/2411.06284


\bibitem{10741141}
Pan, F., Zolfaghari, V., Wen, L., Petrovic, N., Lin, J., Knoll, A.: Generative AI for OCL Constraint Generation: Dataset Collection and LLM Fine-tuning. In: IEEE International Symposium on Systems Engineering (ISSE), pp. 1–8. IEEE (2024). https://doi.org/10.1109/ISSE63315.2024.10741141


\bibitem{Steinberg2008PearsonEducation}
Steinberg, D., Budinsky, F., Paternostro, M., Merks, E.: EMF: Eclipse Modeling Framework. Pearson Education, Boston (2008)

\bibitem{Pilone2005OReillyMedia}
Pilone, D., Pitman, N.: UML 2.0 in a Nutshell. O'Reilly Media, Sebastopol (2005)

\bibitem{llmmde}
Petrovic, N., Fengjunjie, P., Lebioda, K., Zolfaghari, V., Kirchner, S., Purschke, N., Khan, M. A., Vorobev, V., Knoll, A.: Synergy of Large Language Model and Model Driven Engineering for Automated Development of Centralized Vehicular Systems. arXiv preprint arXiv:2404.05508 (2024). https://arxiv.org/abs/2404.05508

\bibitem{Zhang_2024_WACV}
Zhang, G., Zhang, Y., Zhang, K., Tresp, V.: Can Vision-Language Models Be a Good Guesser? Exploring VLMs for Times and Location Reasoning. In: Proceedings of the IEEE/CVF Winter Conference on Applications of Computer Vision (WACV), pp. 636–645 (2024)

\bibitem{zhang2023multimodal}
Zhang, Z., Zhang, A., Li, M., Zhao, H., Karypis, G., Smola, A.: Multimodal Chain-of-Thought Reasoning in Language Models. arXiv preprint arXiv:2302.00923 (2023)

\bibitem{lu2022learn}
Lu, P., Mishra, S., Xia, T., Qiu, L., Chang, K.W., Zhu, S.C., Tafjord, O., Clark, P., Kalyan, A.: Learn to Explain: Multimodal Reasoning via Thought Chains for Science Question Answering. Advances in Neural Information Processing Systems, vol. 35, pp. 2507–2521 (2022)

\bibitem{tan2024retrieval}
Tan, C., Wei, J., Sun, L., Gao, Z., Li, S., Yu, B., Guo, R., Li, S.Z.: Retrieval Meets Reasoning: Even High-School Textbook Knowledge Benefits Multimodal Reasoning. arXiv preprint arXiv:2405.20834 (2024)

\bibitem{wang2022computer}
Wang, S., Zhang, L., Luo, X., Yang, Y., Hu, X., Qin, T., Liu, J.: Computer Science Diagram Understanding with Topology Parsing. ACM Transactions on Knowledge Discovery from Data (TKDD), vol. 16, no. 6, pp. 1–20. ACM New York, NY (2022)

\bibitem{jangra2023survey}
Jangra, A., Mukherjee, S., Jatowt, A., Saha, S., Hasanuzzaman, M.: A Survey on Multi-Modal Summarization. ACM Computing Surveys, vol. 55, no. 13s, pp. 1–36. ACM New York, NY (2023)

\bibitem{che2023enhancing}
Che, C., Lin, Q., Zhao, X., Huang, J., Yu, L.: Enhancing Multimodal Understanding with CLIP-Based Image-to-Text Transformation. In: Proceedings of the 2023 6th International Conference on Big Data Technologies, pp. 414–418 (2023)

\bibitem{ma2022weakly}
Ma, J., Chai, Q., Huang, J., Liu, J., You, Y., Zheng, Q.: Weakly Supervised Learning for Textbook Question Answering. IEEE Transactions on Image Processing, vol. 31, pp. 7378–7388. IEEE (2022)

\bibitem{wang2024cog}
Wang, S., Zhang, L., Zhu, L., Qin, T., Yap, K.H., Zhang, X., Liu, J.: CoG-DQA: Chain-of-Guiding Learning with Large Language Models for Diagram Question Answering. In: Proceedings of the IEEE/CVF Conference on Computer Vision and Pattern Recognition, pp. 13969–13979 (2024)

\bibitem{geipel2024towards}
Geipel, M.M.: Towards a Benchmark of Multimodal Large Language Models for Industrial Engineering. In: 2024 IEEE 29th International Conference on Emerging Technologies and Factory Automation (ETFA), pp. 1–4. IEEE (2024)

\bibitem{openai2024gpt4technicalreport}
OpenAI, Achiam, J., Adler, S., Agarwal, S., Ahmad, L., Akkaya, I., Leoni Aleman, F., et al.: GPT-4 Technical Report. arXiv preprint arXiv:2303.08774 (2024). https://arxiv.org/abs/2303.08774

\bibitem{geminiteam2024gemini15unlockingmultimodal}
Gemini Team, Georgiev, P., Lei, V.I., Burnell, R., Bai, L., Gulati, A., et al.: Gemini 1.5: Unlocking Multimodal Understanding Across Millions of Tokens of Context. arXiv preprint arXiv:2403.05530 (2024). https://arxiv.org/abs/2403.05530

\bibitem{grattafiori2024llama3herdmodels}
Grattafiori, A., Dubey, A., Jauhri, A., Pandey, A., Kadian, A., Al-Dahle, A., et al.: The Llama 3 Herd of Models. arXiv preprint arXiv:2407.21783 (2024). https://arxiv.org/abs/2407.21783

\bibitem{TheClaude3}
Anthropic Inc.: The Claude 3 Model Family: Opus, Sonnet, Haiku. Semantic Scholar, CorpusID: 268232499 (2024). https://api.semanticscholar.org/CorpusID:268232499

\bibitem{agrawal2024pixtral12b}
Agrawal, P., Antoniak, S., Bou Hanna, E., Bout, B., Chaplot, D., Chudnovsky, J., et al.: Pixtral 12B. arXiv preprint arXiv:2410.07073 (2024). https://arxiv.org/abs/2410.07073

\bibitem{xAIGrok}
xAI Inc.: Open Release of Grok-1. xAI Blog (2024). https://x.ai/blog/grok-os

\bibitem{wang2024emu3nexttokenpredictionneed}
Wang, X., Zhang, X., Luo, Z., Sun, Q., Cui, Y., Wang, J., et al.: Emu3: Next-Token Prediction is All You Need. arXiv preprint arXiv:2409.18869 (2024). https://arxiv.org/abs/2409.18869

\bibitem{xue2024xgenmmblip3familyopen}
Xue, L., Shu, M., Awadalla, A., Wang, J., Yan, A., Purushwalkam, S., et al.: xGen-MM (BLIP-3): A Family of Open Large Multimodal Models. arXiv preprint arXiv:2408.08872 (2024). https://arxiv.org/abs/2408.08872

\bibitem{Fuyu8b}
Bavishi, R., Elsen, E., Hawthorne, C., Nye, M., Odena, A., Somani, A., Taşırlar, S.: Fuyu-8B: A Multimodal Architecture for AI Agents. Adept AI Blog (2023). https://www.adept.ai/blog/fuyu-8b

\bibitem{jin2309unified}
Jin, Y., Xu, K., Chen, L., Liao, C., Tan, J., Huang, Q., et al.: Unified Language-Vision Pretraining in LLM with Dynamic Discrete Visual Tokenization. arXiv preprint arXiv:2309.04669 (2024)

\bibitem{peng2023kosmos}
Peng, Z., Wang, W., Dong, L., Hao, Y., Huang, S., Ma, S., Wei, F.: Kosmos-2: Grounding Multimodal Large Language Models to the World. arXiv preprint arXiv:2306.14824 (2023). https://arxiv.org/abs/2306.14824

\bibitem{internvl2}
OpenGVLab: InternVL Family: Closing the Gap to Commercial Multimodal Models with Open-Source Suites —— A Pioneering Open-Source Alternative to GPT-4oOpen Release of Grok-1 (2024). https://github.com/OpenGVLab/InternVL

\bibitem{zhang2024diagramqg}
Zhang, X., Zhang, L., Wu, Y., Huang, M., Wu, W., Li, B., Wang, S., Liu, J.: DiagramQG: A Dataset for Generating Concept-Focused Questions from Diagrams. arXiv preprint arXiv:2411.17771 (2024). https://arxiv.org/abs/2411.17771

\bibitem{wang2021csdqa}
Wang, S., Zhang, L., Yang, Y., Hu, X., Qin, T., Wei, B., Liu, J.: CSDQA: Diagram Question Answering in Computer Science. In: Knowledge Graph and Semantic Computing: Knowledge Graph Empowers New Infrastructure Construction: 6th China Conference, CCKS 2021, Guangzhou, China, November 4-7, 2021, Proceedings, vol. 6, pp. 274–280. Springer (2021)

\bibitem{hiippala2021ai2d}
Hiippala, T., Alikhani, M., Haverinen, J., Kalliokoski, T., Logacheva, E., Orekhova, S., Tuomainen, A., Stone, M., Bateman, J.A.: AI2D-RST: A Multimodal Corpus of 1000 Primary School Science Diagrams. Language Resources and Evaluation, vol. 55, pp. 661–688. Springer (2021)

\bibitem{li2023blip}
Li, J., Li, D., Savarese, S., Hoi, S.: BLIP-2: Bootstrapping Language-Image Pre-Training with Frozen Image Encoders and Large Language Models. In: International Conference on Machine Learning, pp. 19730–19742. PMLR (2023)

\bibitem{radford2021learningtransferablevisualmodels}
Radford, A., Kim, J.W., Hallacy, C., Ramesh, A., Goh, G., Agarwal, S., Sastry, G., Askell, A., Mishkin, P., Clark, J., Krueger, G., Sutskever, I.: Learning Transferable Visual Models From Natural Language Supervision. arXiv preprint arXiv:2103.00020 (2021). https://arxiv.org/abs/2103.00020

\bibitem{conrardy2024image}
Conrardy, A., Cabot, J.: From Image to UML: First Results of Image-Based UML Diagram Generation Using LLMs. arXiv preprint arXiv:2404.11376 (2024). https://arxiv.org/abs/2404.11376

\bibitem{rossi2024importance}
Rossi, R.: The Importance of Visual Modelling Languages in Generative Software Engineering. arXiv

\bibitem{llama38b}
HuggingFace:Meta-Llama-3-8B-Instruct.https://huggingface.co/meta-llama/Meta-Llama-3-8B-Instruct

\bibitem{internvl28b}
HuggingFace: OpenGVLab/InternVL2-8B. https://huggingface.co/OpenGVLab/InternVL2-8B


\bibitem{quart}
Pallets Project: Quart. https://quart.palletsprojects.com/en/latest/

\bibitem{seo2014diagram}
Seo, M. J., Hajishirzi, H., Farhadi, A., Etzioni, O.: Diagram understanding in geometry questions. In: Proceedings of the AAAI Conference on Artificial Intelligence, vol. 28, no. 1 (2014).

\bibitem{zhang2022plane}
Zhang, M.-L., Yin, F., Hao, Y.-H., Liu, C.-L.: Plane geometry diagram parsing. arXiv preprint arXiv:2205.09363 (2022). https://arxiv.org/abs/2205.09363

\bibitem{saikh2022scienceqa}
Saikh, T., Ghosal, T., Mittal, A., Ekbal, A., Bhattacharyya, P.: ScienceQA: A novel resource for question answering on scholarly articles. International Journal on Digital Libraries, vol. 23, no. 3, pp. 289--301 (2022). Springer.

\bibitem{schwenk2022okvqa}
Schwenk, D., Khandelwal, A., Clark, C., Marino, K., Mottaghi, R.: A-OKVQA: A benchmark for visual question answering using world knowledge. In: European Conference on Computer Vision, pp. 146--162 (2022). Springer.

\bibitem{ma2023xtqa}
Ma, J., Chai, Q., Liu, J., Yin, Q., Wang, P., Zheng, Q.: XTQA: Span-level Explanations for Textbook Question Answering. *IEEE Transactions on Neural Networks and Learning Systems* (2023). IEEE.


\bibitem{wu2020comprehensive}
Wu, Z., Pan, S., Chen, F., Long, G., Zhang, C., Yu, S.: A Comprehensive Survey on Graph Neural Networks. *IEEE Transactions on Neural Networks and Learning Systems*, **32**(1), 4--24 (2020). IEEE.

\bibitem{2023lmdeploy}
LMDeploy Contributors: LMDeploy: A Toolkit for Compressing, Deploying, and Serving LLM. GitHub repository (2023). Available: https://github.com/InternLM/lmdeploy


\end{thebibliography}
\end{document}